\documentclass[a4paper,reqno]{amsart}
\usepackage{amsmath}
\usepackage{amstext}
\usepackage{amsfonts}
\usepackage{amssymb}
\usepackage{amscd}
\usepackage[foot]{amsaddr}
\usepackage{natbib}
\usepackage{graphicx}
\usepackage{placeins}
\usepackage{appendix}

\newtheorem{theorem}{Theorem}

\newtheorem{conjecture}[theorem]{Conjecture}
\newtheorem{corollary}[theorem]{Corollary}

\newtheorem{notation}[theorem]{Notation}

\newtheorem{remark}[theorem]{Remark}

\graphicspath{{d:/artykuly/dHondt/}}
\input{tcilatex}
\DeclareMathOperator*{\E}{\mathbb{E}}

\calclayout

\begin{document}

\title[The Number of Parties and the District Magnitude]{A Formal Model of the Relationship between the Number of Parties and
the District Magnitude}

\author[D. Boratyn]{Daria Boratyn $^1,^2$}
\email{daria.boratyn@im.uj.edu.pl}

\author[J. Flis]{Jaros\l aw Flis $^1,^3$}
\email{jaroslaw.flis@uj.edu.pl}

\author[W. S\l omczy\'{n}ski]{Wojciech S\l omczy\'{n}ski $^1,^2$}
\email{wojciech.slomczynski@im.uj.edu.pl}

\author[D. Stolicki]{Dariusz Stolicki $^1,^4$}
\email{dariusz.stolicki@uj.edu.pl}

\address{$^1$ Jagiellonian Center for Quantitative Research in Political Science \newline
	Jagiellonian University, ul. Wenecja 2, 31-117 Krak\'{o}w, Poland}
\address{$^2$ Institute of Mathematics, \newline
	Jagiellonian University, ul. \L{}ojasiewicza 6, 30-348 Krak\'{o}w, Poland}
\address{$^3$ Institute of Journalism, Media, and Social Communications, \newline Jagiellonian University, ul. \L{}ojasiewicza 4, 30-348 Krak\'{o}w, Poland}
\address{$^4$ Institute of Political Science and International Relations, \newline Jagiellonian University, ul. Jab\l{}onowskich 5, 31-114 Krak\'{o}w, Poland}

\date{\today}

\maketitle

\begin{abstract}
On the basis of a formula for calculating seat shares and natural thresholds
in multidistrict elections under the Jefferson-D'Hondt system and a
probabilistic model of electoral behavior based on P\'{o}lya's urn model, we
propose a new model of the relationship between the district magnitude and
the number / effective number of relevant parties. We test that model on
both electoral results from multiple countries employing the D'Hondt method
(relatively small number of elections, but wide diversity of
political configurations) and data based on hundreds of Polish local
elections (large number of elections, but much higher degree
of parameter uniformity). We also explore some applications of the proposed
model, demonstrating how it can be used to estimate the potential effects of
electoral engineering.
\end{abstract}

\section{Introduction}

The study of the relationship between the number of relevant (i.e.
seat-winning) parties and the district magnitude dates back at least to the
1950's and the celebrated \emph{Duverger's law} 
\citep{Duverger51, Duverger54}
claiming that two-party systems tend to be correlated with the use of
electoral plurality rule (FPTP). Subsequent work by 
\citet{Rae67}%
, 
\citet{TaageperaLaakso80}%
, 
\citet{TaageperaShugart89}%
, 
\citet{Lijphart90}%
, and 
\citet{Cox97}
have led to a generalization of Duverger's law stating that the number of
relevant parties is correlated with the district magnitude, or, more
precisely, that it is an increasing and concave function thereof. Its
classic formulation is the well-known \emph{micromega rule}\textbf{\ }%
\citep{Colomer04}%
. The relationship thus revealed is due to a combination of three effects 
\citep[103-104]{Taagepera07}%
: the \emph{mechanical effect}, arising from the operation of the seat
apportionment formula, and the \emph{psychological effects} both on parties and
voters. Parties, seeking to maximize their chance of being relevant, adjust
to small district magnitudes by the process of integration, leading to a
smaller number of contending parties, while voters, afraid of wasting their
votes, vote strategically for larger parties, thus affecting the
distribution of vote shares.

There is wealth of recent empirical evidence supporting the above findings.
See, e.g., 
\citet{Ziegfeld13}%
, 
\citet{BarceloMuraoka18}%
, 
\citet{SingerGershman18}%
. There are, however, relatively few formal (as opposed to purely
statistical) models of the relationship between the number of relevant
parties and the district magnitude. Of those, the most prominent is the
Seat-Product Model, introduced by 
\citet{TaageperaShugart93}%
, and further developed by 
\citet{Taagepera07}%
, 
\citet{LiShugart16}%
, and 
\citet{ShugartTaagepera17}%
. It can be expressed as follows:

\begin{conjecture}[Seat-Product Model]
The number of relevant parties $n$ can be approximated by%
\begin{equation}
n\approx n_{SPM}:=\left( ms\right) ^{1/4},
\end{equation}%
where $m$ is the mean district magnitude and $s$ is the assembly size (i.e.,
the total number of seats).
\end{conjecture}

Taagepera's and Shugart's reasoning is essentially as follows: first, the
distribution of the number of relevant parties in a single district of
magnitude $m_{d}$ is approximated by an absolutely continuous distribution
with support bounded by the two logical extremes, i.e., $n_{d}=1$ and $%
n_{d}=m_{d}$. Then it is noted that said distribution is right-skewed. From
this fact it is assumed, by a reasoning loosely analogous to the well-known
principle of insufficient reason, that the expectation of that distribution
should be equal to the geometric mean of those two bounds, i.e., $m_{d}^{1/2}
$. It is then noted that distribution of the number of relevant parties
nationwide can likewise be approximated by an absolutely continuous
distribution with support again bounded by the two logical extremes, i.e., $%
n=m_{d}$ (since parties relevant in a single districts are always nationally
relevant) and $n=s^{1/2}$. Again by assuming that the expectation should be
equal to the geometric mean of the two, we arrive at $n=\left( ms\right)
^{1/4}$, as above.

While certainly attractive in its simplicity, the Seat-Product Model has a
number of weaknesses. Being only a function of the assembly size and the
district magnitude, and thereby failing to account for such parameters as
the apportionment formula, electoral thresholds, or sociopolitical
cleavages, its empirical accuracy is necessarily limited. At the same time,
the theoretical justification for the Seat-Product Model is highly informal.
It would appear that its authors' intended to provide a formula for expected
number of relevant parties, but if that is the case, the underlying
probabilistic assumptions are not expressly specified.

We propose an alternative model of the expected number of relevant parties
in electoral systems employing the Jefferson-D'Hondt formula for
intra-district seat apportionment (we note, however, that our model can be
extended to other PR\ formulae, although formulation and testing of such
extensions is beyond the scope of this article). It is based on the
\textquotedblleft pot and ladle\textquotedblright\ model of the
Jefferson-D'Hondt method, formulated by three of us 
\citet{FlisEtAl19}
in a recent article, and on a result by two of us (Boratyn and Stolicki)
regarding the distribution of the vote shares of the $k$-th largest party if
the election result is drawn from the unit simplex with a uniform
distribution. The two main advantages of our formula over the Seat-Product
Model lie in its improved accuracy (as demonstrated by the empirical tests
below) and its more formal theoretical justification. At the same time, it
is more limited in scope: it only applies to electoral systems employing the
Jefferson-D'Hondt formula (although that limitation becomes less restrictive
as one realizes that FPTP is just a limiting case of Jefferson-D'Hondt),
involves an additional parameter in the form of the number of registered
parties, is limited to the mechanical effects, and assumes that the
psychological effect on voters is negligible (as otherwise the assumption of
uniform distribution of electoral results would be less defensible).

\begin{remark}
Note that we do not address the question of the psychological effect on
parties. Ideally, if such effect were to be assumed, it should be modeled
before the mechanical effects in order to estimate the number of registered
parties, which appears as a parameter in our formula. Here, we only note
that our empirical data is inconclusive with regard to the existence of the
psychological effect: Nadaraya-Watson kernel regression, employed due to the
fact that we have no theoretical model for the functional form of the
suspected relationship between mean district magnitude and the number of
relevant parties, reveals no regular relationship between the two for Polish
county elections dataset, and is wholly inconclusive for the European
national elections dataset because of the limited number of points.
\end{remark}

\begin{figure}[h]
	\centering
	\includegraphics[width=0.8\linewidth]{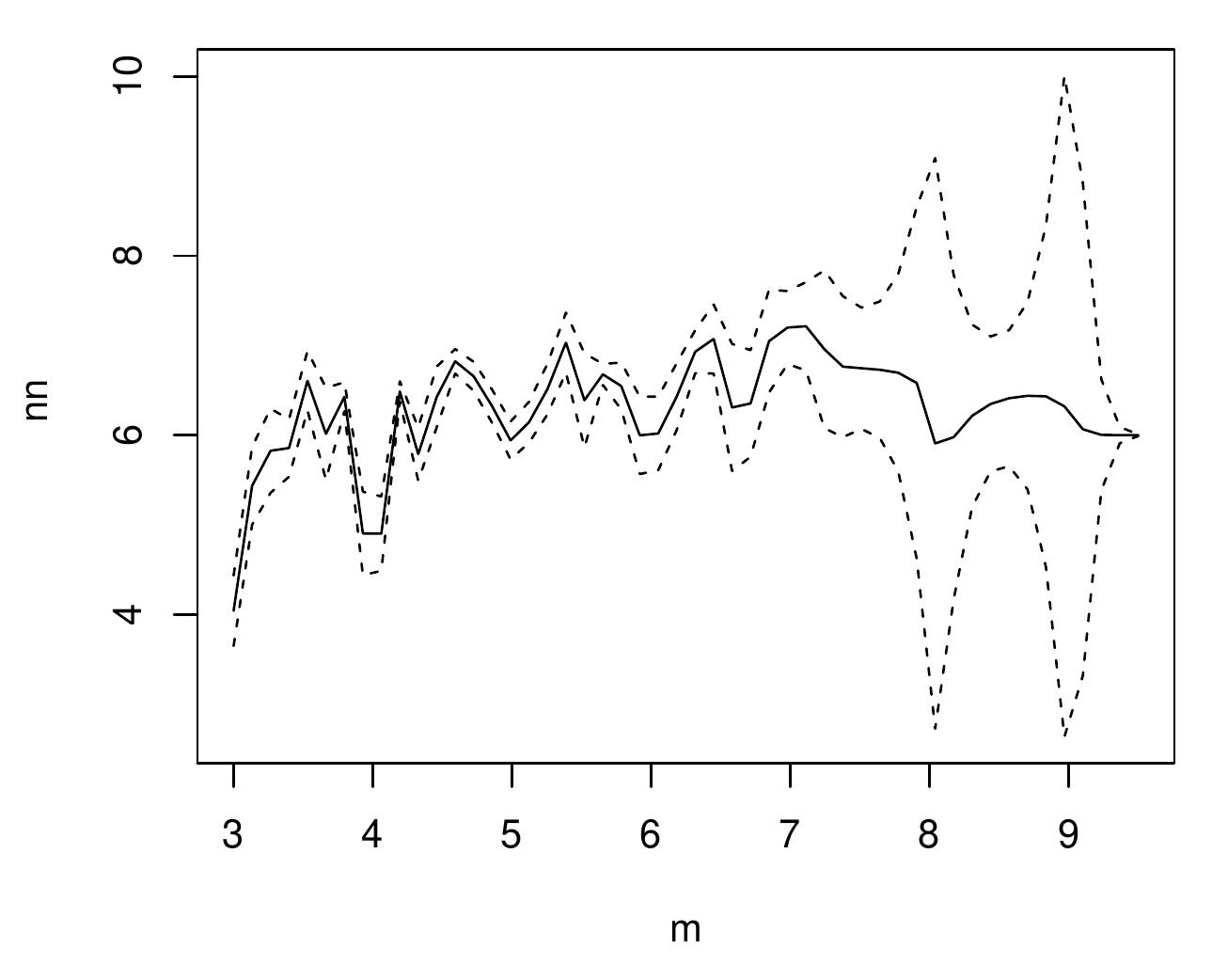}
	\caption{Kernel regression estimate (with 95		between the mean district magnitude and the number of registered parties in Polish county elections.}
	\label{fig:npreg-nn-m}
\end{figure}%

We begin with setting up the notation to be used throughout the remainder of
this article.

\begin{notation}
Let:

\begin{itemize}
\item $x_{i}^{\downarrow }$ denote the $i$\textbf{-th largest element} of a
totally ordered countable set $\{x_{1},x_{2},\ldots \}$,

\item $s\in \mathbb{N}_{+}$ be the \textbf{assembly size}, i.e., total
number of seats,

\item $c\in \mathbb{N}_{+}$ be the \textbf{number of districts},

\item $N\in \mathbb{N}_{+}$ be the \textbf{number of registered parties},

\item $n\in \left\{ 1,\dots ,N\right\} $ be the \textbf{number of relevant
parties},

\item $m:=s/c$ be the \textbf{mean district magnitude},

\item $\Delta _{n}$ be the $\left( n-1\right) $\textbf{-dimensional
probability simplex},

\item $\mathbf{v}:=\left( v_{1},\dots ,v_{N}\right) \in \Delta _{n}$ be the 
\textbf{nationwide vote share vector},

\item $p_{i}:=v_{i}/\sum_{j=1}^{n}v_{j}^{\downarrow }$ be the nationwide
vote share of the $i$-th relevant party, $i=1,\dots ,n$, \textbf{renormalized%
} after excluding \textbf{non-relevant parties},

\item $q_{i}^{\downarrow }:=v_{i}^{\downarrow
}/\sum_{j=1}^{i}v_{j}^{\downarrow }$ be the nationwide vote share of the $i$%
-th largest party, $i=1,\dots ,N$, \textbf{renormalized} after excluding 
\textbf{all smaller parties},

\item $v_{i}^{k}$ be the \textbf{vote share} of the $i$-th party \textbf{in
the }$k$\textbf{-th district}, $k=1,\dots ,c$,

\item $p_{i}^{k}$ be the vote share of the $i$-th relevant party in the $k$%
-th district renormalized after excluding non-relevant parties,

\item $M_{k}\in \mathbb{R}_{+}$ be the \textbf{Jefferson-D'Hondt multiplier}
in the $k$-th district,

\item $R_{i}^{k}:=\left\{ p_{i}^{k}M_{k}\right\} $ be the \textbf{rounding
residual} of the $i$-th relevant party in the $k$-th district,

\item $\tau \in \left( 0,1\right) $ be the \textbf{statutory threshold},

\item $\left\langle x_{i}^{k}\right\rangle $ denote the \textbf{average} of $%
x_{i}^{k}$ over $k=1,\dots ,c$.
\end{itemize}
\end{notation}

\section{Theoretical Model}

\subsection{The \textquotedblleft Pot and Ladle\textquotedblright\ Model of
the Jefferson-D'Hondt System}

The starting point of our reasoning consists of the \textquotedblleft pot
and ladle\textquotedblright\ model of the Jefferson-D'Hondt system,
introduced in 
\citep{FlisEtAl19}
and further elaborated 
\citep{FlisEtAl18}%
, which provides the following closed-form formula for estimating seat
counts from nationwide vote shares:

\begin{theorem}
\label{thmPotAndLadle}Assume there are $n$ relevant parties. If

\begin{enumerate}
\item[(A1)] there exists a selection of multipliers $M_{1},\dots ,M_{k}$
such that for each $j=1,\dots ,n$

\begin{enumerate}
\item[(A1a)] rounding residuals average to $1/2$ over districts, i.e., $%
\left\langle R_{j}^{k}\left( M_{k}\right) \right\rangle =\frac{1}{2}$;

\item[(A1b)] multipliers are not correlated with normalized vote shares,
i.e.,\linebreak $\func{Cov}\left( p_{j}^{k},M_{k}\right) =0$;
\end{enumerate}

\item[(A2)] no seats have been obtained by non-relevant parties;

\item[(A3)] for each $j=1,\dots ,n$ the normalized nationwide vote share
equals average normalized district-level vote share, i.e., $%
p_{j}=\left\langle p_{j}^{k}\right\rangle $,
\end{enumerate}

then for each relevant party, $j=1,\dots ,n$, the nationwide number of seats
is given by%
\begin{equation}
s_{j}=p_{j}s+p_{j}\frac{cn}{2}-\frac{c}{2}.  \label{seatAlloc}
\end{equation}
\end{theorem}

For proof of the above, see 
\citep{FlisEtAl18}%
. Of the above assumptions, \textbf{A2} and \textbf{A3} are essentially
political in nature, requiring that there be no anomalies in the spatial
distribution of party support (such as regionalisms, malapportionment, or
large differences in district magnitudes) that cause district-level
electoral results to diverge from nationwide data. In constrast, \textbf{A1}
is rather technical, but can be justified mathematically: under some general
distributional assumptions\footnote{%
District magnitudes $S_{1},\dots ,S_{c}$ are independent random variables
identically distributed according to some discrete probability distribution
on $\mathbb{N}_{+}$ with expectation $m\in \mathbb{[}S_{\min },\infty 
\mathbb{)}$, where $S_{\min }\geq 1$; district-level vote share vectors $%
\mathbf{P}_{1},\dots ,\mathbf{P}_{c}$ are independent random variables
identically distributed according to some absolutely continuous probability
distribution on the $(n-1)$-dimensional unit simplex $\Delta _{n}$ with
expectation $\mathbf{p}:=\left( p_{1},\dots ,p_{n}\right) \in \Delta _{n}$
and some continously differentiable density vanishing at the faces of $%
\Delta _{n}$ and non-increasing within radius $\left( S_{\min }+1\right)
^{-1}$ from the vertices of $\Delta _{n}$.}, it can be shown that there
exist multipliers for which both \textbf{A1a} and \textbf{A1b} hold
approximately, and thus the expected number of seats $\func{E}\left(
s_{j}\right) $ is well approximated by (\ref{seatAlloc}). Empirical research
shows that both sets of assumptions correspond closely to reality, making
the \textquotedblleft pot-and-ladle\textquotedblright\ model quite accurate.

One of the advantages of the \textquotedblleft pot and
ladle\textquotedblright\ model from our point of view consists of the fact
that it includes a mathematically convenient criterion of relevance:

\begin{corollary}
Assuming there are no statutory thresholds, the $j$-th party is relevant in
the sense of Theorem \ref{thmPotAndLadle} if and only if%
\begin{equation}
p_{j}\geq t:=\frac{1}{2m+n}.  \label{naturalThold}
\end{equation}
\end{corollary}

In the above formula, $t$ denotes the \emph{natural threshold}\footnote{%
The concept of the \emph{natural threshold of inclusion}, i.e., the minimal
vote share necessary to obtain a non-zero number of seats, has been first
introduced by 
\citet{Rokkan68}%
. 
\citet{RaeEtAl71}
have later introduced a complementary concept of the \emph{threshold of
exclusion}, i.e., maximum vote share under which a party can fail to obtain
a non-zero number of seats. Natural thresholds for diverse methods of seat
apportionment have been obtained in the following years by, \textit{inter
alia}, 
\citet{LoosemoreHanby71}%
, 
\citet{LijphartGibberd77}%
, 
\citet{Laakso79a}%
, and 
\citet{PalomaresRamirezGonzalez03}%
. Unfortunately, those theoretical results are applicable only within a
single electoral district. Nationwide thresholds cannot be estimated
precisely without additional assumptions regarding the distribution of party
support over districts, although several widely applicable heuristics have
been proposed by 
\citet{Taagepera89,Taagepera98,Taagepera98a,Taagepera02}%
.}. The underlying reasoning is quite simple: for $p_{j}<t$, assumptions 
\textbf{A1} and \textbf{A3} lead to a contradiction. But an apparent
circularity in reasoning can be noted: the natural threshold $t$ depends on
the number of relevant parties, which in turn depends on $n$. This
circularity can be eliminated when one recognizes that the above Corollary
is equivalent to:

\begin{corollary}
Assuming there are no statutory thresholds, the $k$-th largest party is
relevant in the sense of Theorem \ref{thmPotAndLadle} if and only if%
\begin{equation}
p_{k}^{\downarrow }\geq t_{k}:=\frac{1}{2m+k}.  \label{naturalTholdK}
\end{equation}
\end{corollary}

In systems employing statutory thresholds as well, there is of course a
concurrent criterion that%
\begin{equation}
v_{k}^{\downarrow }\geq \tau .  \label{statThold}
\end{equation}

It further follows that:

\begin{corollary}
The number of relevant parties appearing in (\ref{seatAlloc}) is given by%
\begin{equation}
n:=\max \left\{ k=1,\dots ,N:p_{k}^{\downarrow }\geq \frac{1}{2m+k}~\text{and%
}~v_{k}^{\downarrow }\geq \tau \right\} .  \label{relevParties}
\end{equation}
\end{corollary}

\subsection{Modeling the Distribution of Party Vote Shares}

What is the probability that the $k$-th largest party is relevant under Eq.~(%
\ref{naturalTholdK}) and under Eq.~(\ref{statThold})? It is clear that the
answer depends on the distribution of votes among parties (note that unlike
in the preceding chapter, the random variable here is the nationwide rather
than district-level vector of vote shares). The problem of modeling such
distribution is equivalent to a special case of the problem of modeling
preference orderings, which is well known in the social choice theory (see,
e.g., 
\citealp{RegenwetterEtAl06}
and 
\citealp{TidemanPlassmann12}%
). Of those, motivated by the principle of insufficient reason, we choose
the \emph{Impartial Anonymous Culture (IAC)} model which treats each
preference profile (and, thus, each distribution of votes among parties) as
equiprobable 
\citealp{GehrleinFishburn76a,KugaNagatani74}%
\footnote{%
The assumption of uniform distribution of party vote shares, which underlies
the IAC model, has been made by a number of authors working on mathematical
voting theory, including 
\citealt{Happacher01,SchusterEtAl03,SchwingenschloglDrton04,JansonLinusson11}%
; and others.}. Accordingly, the vote share vector $\mathbf{v}$ follows the
uniform distribution on a~discrete grid of points within $\Delta _{N}$,
which, as the number of voters approaches infinity, weakly converges to the
uniform distribution on $\Delta _{N}$. To simplify calculations, we focus on
the limiting case.

Under those distributional assumptions, we arrive at two results regarding
the distribution of absolute and normalized vote shares of the $k$-th
largest party:

\begin{theorem}
\label{thm:kPlayer}If $\mathbf{V\sim }\func{Unif}\left( \Delta _{N}\right) $%
, where $\func{Unif}\left( \Delta _{N}\right) $ is the uniform distribution
on $\Delta _{N}$, then $V_{k}^{\downarrow }$, $k=1,\dots ,N$, is distributed
according to a continuous distribution supported on $[0,1/k]$ for $k>1$ and
on $[1/N,1]$ for $k=1$, with density $f_{V_{k}^{\downarrow }}$ given by:%
\begin{equation}
f_{V_{k}^{\downarrow }}\left( x\right) :=N\left( N-1\right) \dbinom{N-1}{k-1}%
\sum\limits_{a=k}^{\min \left( N,\left\lfloor 1/x\right\rfloor \right)
}\left( -1\right) ^{a-k}\dbinom{N-k}{a-k}\left( 1-ax\right) ^{N-2}.
\label{densKthPlayer}
\end{equation}%
\begin{figure}[h]
	\centering
	\includegraphics[width=0.65\linewidth]{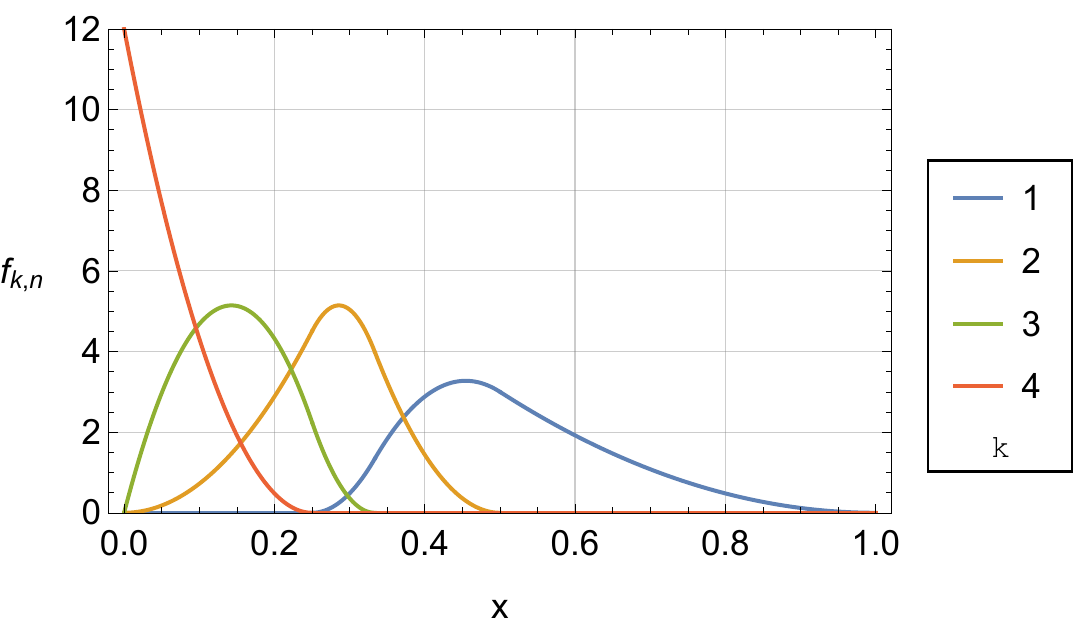}
	\caption{Density of the distribution of the voting weight of the $k$-th largest of $4$ players.}
	\label{fig:kDensity}
\end{figure}%
\end{theorem}

\begin{theorem}
\label{thm:kPlayerNorm}If $\mathbf{V\sim }\func{Unif}\left( \Delta
_{N}\right) $, where $\func{Unif}\left( \Delta _{N}\right) $ is the uniform
distribution on $\Delta _{N}$, then $P_{1}^{\downarrow }=1$ with probability 
$1$, and $P_{k}^{\downarrow }$, $k=2,\dots ,N$, is distributed according to
a continuous distribution supported on $[0,1/k]$, with density $%
f_{P_{k}^{\downarrow }}$ given by:%
\begin{multline}
f_{P_{k}^{\downarrow }}\left( x\right) :=\mathbf{1}_{\left\{ z<1/k\right\}
}\,k\left( k-1\right) \dbinom{N}{k}\left( 1-xk\right) ^{k-2}\times 
\label{densNormKthPlayer} \\
\times \sum_{a=k}^{N}\left( -1\right) ^{a-k}\dbinom{N-k}{a-k}\left( 1+\left(
a-k\right) x\right) ^{-k}.
\end{multline}
\end{theorem}

For proofs of the above theorems, see Appendix A%
\nocite{BoratynEtAl19}%
.

Note that Theorem \ref{thm:kPlayer} corresponds to a known earlier result on
the order statistics of uniform spacings 
\citep{Darling53,RaoSobel80,Devroye81}%
. However, it is impossible to obtain Theorem \ref{thm:kPlayerNorm} from
that result alone, and proof techniques employed in the context of uniform
spacings are not easily extended to that question. In contrast, our
(original) proof of Theorem \ref{thm:kPlayer}, given in the Appendix A,
provides a foundation of the proof of Theorem \ref{thm:kPlayerNorm} as well.

\subsection{Probability of Relevance}

From Theorems \ref{thm:kPlayerNorm} and \ref{thm:kPlayerNorm} we can obtain
a closed-form formula for the probability of the $k$-th largest party being
relevant. Recall from (\ref{naturalTholdK}) that in the absence of a
statutory threshold, the $k$-th largest party is relevant if and only if%
\begin{equation}
p_{k}^{\downarrow }\geq t_{k}:=\left( 2m+k\right) ^{-1}.
\end{equation}%
Thus, we are interested in $\Pr \left( P_{k}^{\downarrow }>t_{k}\right) $.
By integrating the density of $P_{k}^{\downarrow }$, we obtain (see Appendix
B for details):%
\begin{equation}
\Pr \left( P_{k}^{\downarrow }>t_{k}\right) =k\dbinom{N}{k}%
\sum_{a=k}^{N}\left( -1\right) ^{a-k}\dbinom{N-k}{a-k}\frac{1}{a}\left( 
\frac{2m}{a+2m}\right) ^{k-1}.  \label{probRelevance}
\end{equation}

Likewise, the $k$-th largest party is relevant under a statutory threshold
if $v_{k}^{\downarrow }\geq \tau .$By integrating the density of $%
V_{k}^{\downarrow }$, we obtain:%
\begin{equation}
\Pr \left( V_{k}^{\downarrow }>\tau \right) =N\dbinom{N-1}{k-1}%
\sum_{a=k}^{\min \left( N,\left\lfloor \tau ^{-1}\right\rfloor \right)
}\left( -1\right) ^{a-k}\dbinom{N-k}{a-k}\frac{1}{a}\left( 1-a\tau \right)
^{N-1}.  \label{probRelevST}
\end{equation}

\subsection{Expected Number of Relevant Parties}

Expected number of relevant parties is easy to obtain from the above
results. First, let us note that the probability that the $k$-th party is
last relevant party can be expressed as the difference of the probabilities
of relevance of the $k$-th and $\left( k+1\right) $-th parties (with $\Pr
\left( P_{N+1}^{\downarrow }>t_{N+1}\right) =\Pr \left( V_{N+1}^{\downarrow
}>\tau \right) =0$):%
\begin{equation}
\Pr \left( n=k\right) =\Pr \left( P_{k}^{\downarrow }>t_{k}\right) -\Pr
\left( P_{k+1}^{\downarrow }>t_{k+1}\right) .
\end{equation}%
Now from the formula for the expectation of a discrete distribution we have
(see Appendix C for details):%
\begin{equation}
\func{E}n=\sum_{k=1}^{N}k\Pr \left( n=k\right) =\sum_{k=1}^{N}\Pr \left(
P_{k}^{\downarrow }>t_{k}\right) \text{,}  \label{expectRelev}
\end{equation}%
and if statutory thresholds are accounted for,%
\begin{equation}
\func{E}n=\min \left\{ \sum_{k=1}^{N}\Pr \left( P_{k}^{\downarrow
}>t_{k}\right) ,\sum_{k=1}^{N}\Pr \left( V_{k}^{\downarrow }>\tau \right)
\right\} .  \label{expectRelevST}
\end{equation}

We can likewise obtain the formula for the second moment of $n$,%
\begin{equation}
\func{E}\left( n^{2}\right) =\sum_{k=1}^{N}k^{2}\Pr \left( n=k\right)
=2\sum_{k=2}^{N}k\Pr \left( P_{k}^{\downarrow }>t_{k}\right)
-\sum_{k=1}^{N}\Pr \left( P_{k}^{\downarrow }>t_{k}\right) \text{,}
\end{equation}%
and thus for variance,%
\begin{gather}
\func{Var}\left( n\right) =\func{E}\left( n^{2}\right) -\left( \func{E}%
n\right) ^{2}=  \notag \\
=2\sum_{k=2}^{N}k\Pr \left( P_{k}^{\downarrow }>t_{k}\right)
-\sum_{k=1}^{N}\Pr \left( P_{k}^{\downarrow }>t_{k}\right) -\left(
\sum_{k=1}^{N}\Pr \left( P_{k}^{\downarrow }>t_{k}\right) \right) ^{2}.
\end{gather}

\subsection{Plots}

The behavior of formulae (\ref{probRelevance}) and (\ref{expectRelev}) is
illustrated by the following plots.

\begin{figure}[h]
	\centering
	\includegraphics[width=0.475\linewidth]{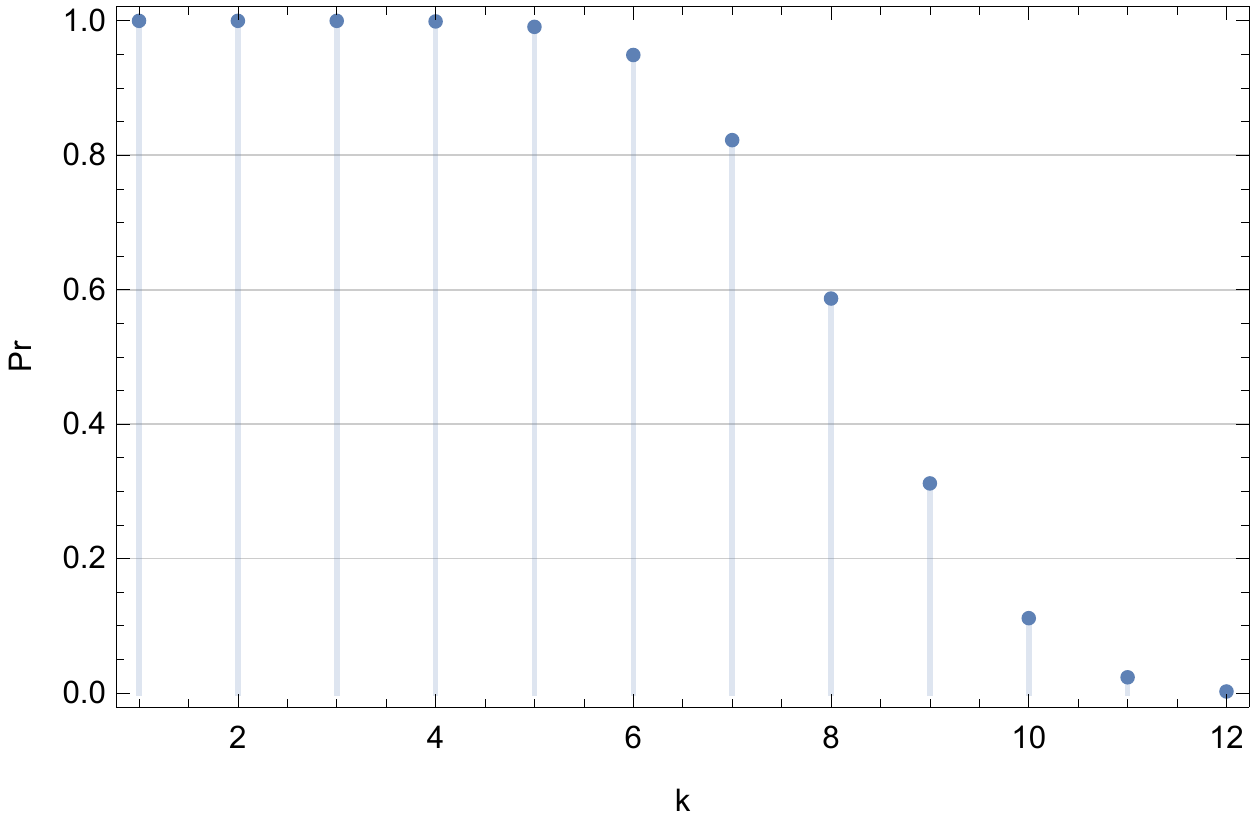}
	\includegraphics[width=0.475\linewidth]{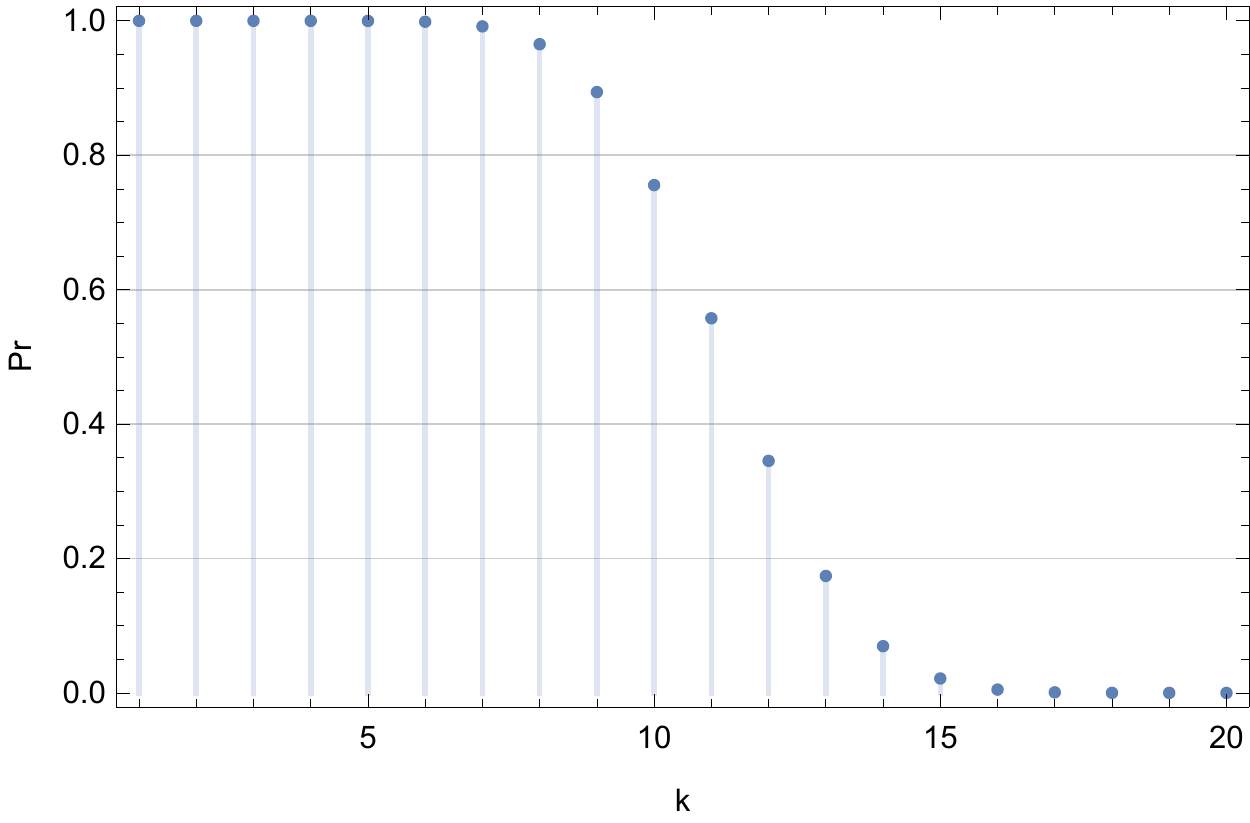}
	\caption{Probability of relevance of the $k$-th party, $k = 1, \dots, N$, out of
		$N = 12$ (left) or $N = 20$ (right) assuming mean district magnitude $m = 8$.}
	\label{fig:probRelev-k}
\end{figure}%

\begin{figure}[h]
	\centering
	\includegraphics[width=0.6\linewidth]{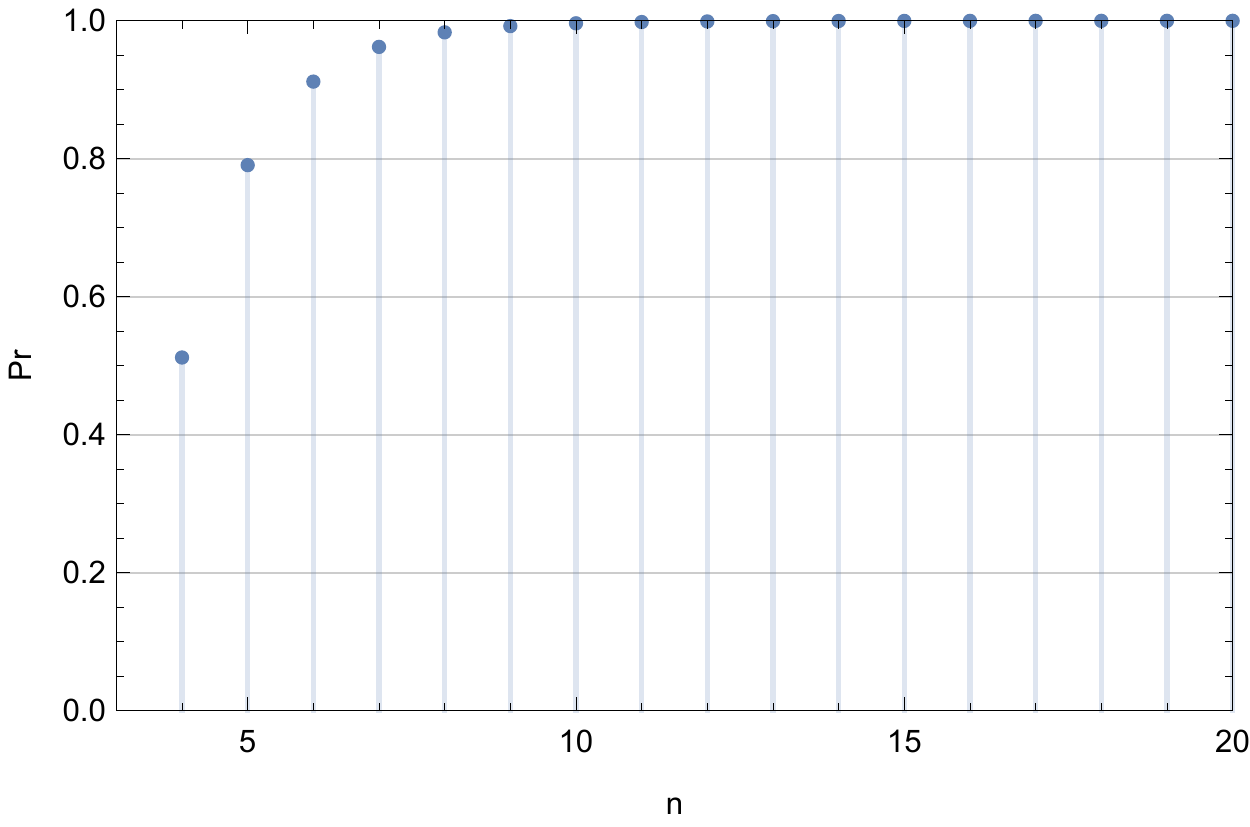}
	\caption{Probability of relevance of the $4$-th party out of $N = 4, \dots, 20$, assuming
		mean district magnitude $m = 8$.}
	\label{fig:probRelev-n-k4}
\end{figure}%

\begin{figure}[h]
	\centering
	\includegraphics[width=0.8\linewidth]{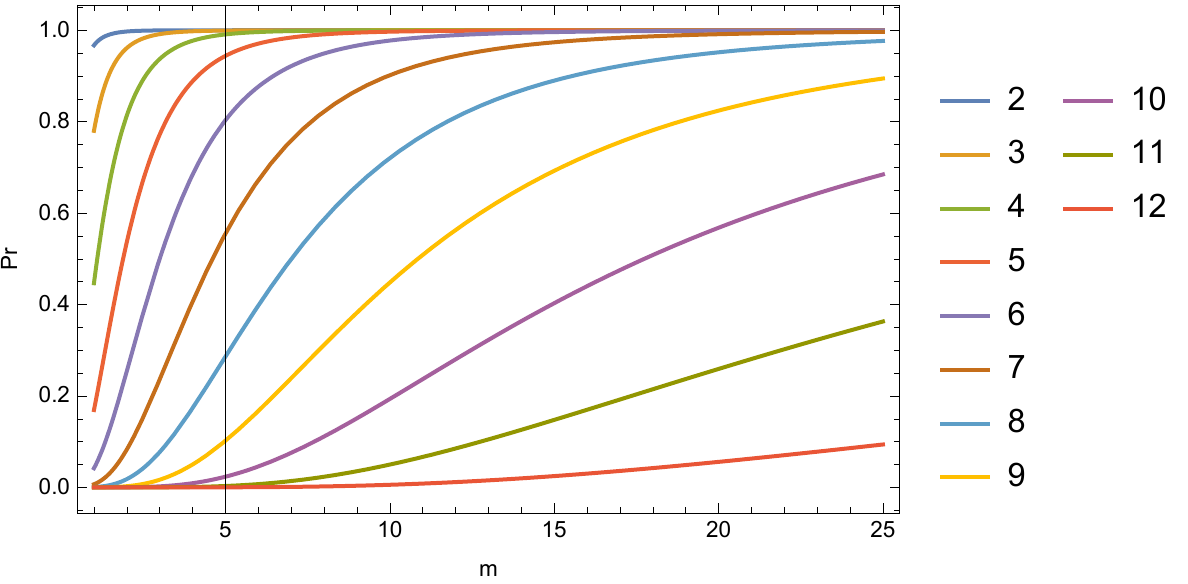}
	\caption{Probability of relevance of the $k$-th party out of $N = 12$ as a function of mean district magnitude $m$.}
	\label{fig:probRelev-m}
\end{figure}%

\begin{figure}[h]
	\centering
	\includegraphics[width=0.8\linewidth]{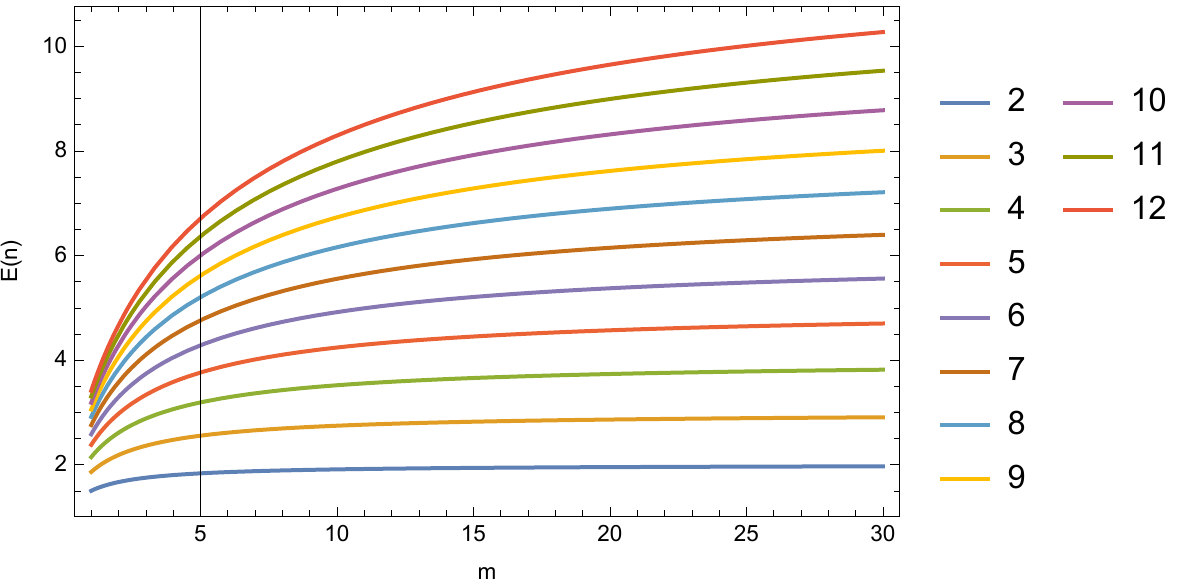}
	\caption{Expected number of relevant parties as a function of mean district magnitude $m$ and the number
		of registered parties $N = 2, \dots, 12$.}
	\label{fig:expRelev-m}
\end{figure}%

\FloatBarrier%

\section{Empirical Test}

We have tested our model using two sets of data: general elections results
from nine EU countries using the Jefferson--D'Hondt method (with Belgium
split into the Flemish and Waloon regions due to the disjointness of their
party systems) and Polish county council elections. The European general
election dataset is rather small (87 elections), but highly heterogenous in
terms of both parameters of interest (i.e., the district magnitude varies
between $6.731$ and $150$, and the number of registered parties varies
between $3$ and $28$). The Polish county election dataset is more numerous
(1564 elections), but more homogenous ($m$ varies between $3$ and $9.5$, and 
$N$ -- between $2$ and $15$).

\subsection{General Elections in Nine EU Countries Using Jefferson-D'Hondt}

There are currently nine EU member states using the Jefferson--D'Hondt
method for parliamentary seat allocation in multimember districts (we
exclude countries such as Austria that only use JDH in combination with
other methods, e.g., in multi-tiered elections): Belgium, Croatia, the Czech
Republic, Finland, Luxembourg, Netherlands, Poland, Portugal, and Spain. For
each of those nine countries, we analyze all post-1945 elections held under
JDH rule (as noted above, we treat the Belgian general election of 2014, the
only one held under pure JDH, as a special case, i.e., as two concurrent but
distinct elections in Flemish and Walloon regions). For each election, we
have identified the mean district magnitude $m$ and the number of registered
national parties (defining a registered national party as a party or a
single-list coalition\ fielding candidates in at least 90\% of electoral
districts), computed the expected number of relevant parties using our
probabilistic model, and compared the outcome with the one arising under the
Seat-Product Model. Table \ref{tbl:intlErrors} gives mean $L_{1}$ errors
(absolute deviations from the mean) for our model, the Seat-Product Model,
and \textquotedblleft naive\textquotedblright\ $n=N$ model.

\begin{center}
\begin{table}[h] \centering%
\begin{tabular}{c|ccc}
country & probabilistic & SPM & $n=N$ \\ \hline\hline
Belgium -- F & 0.128 & 0.039 & 0.333 \\ 
Belgium -- W & 0.051 & 0.276 & 0.428 \\ 
Croatia & 0.226 & 0.238 & 1.988 \\ 
Czech Republic & 0.368 & 0.397 & 2.002 \\ 
Finland & 0.125 & 0.235 & 0.530 \\ 
Luxembourg & 0.112 & 0.163 & 0.146 \\ 
Netherlands & 0.750 & 0.240 & 1.014 \\ 
Poland & 0.083 & 0.433 & 0.792 \\ 
Portugal & 0.433 & 0.495 & 1.096 \\ 
Spain & 0.328 & 0.378 & 0.848%
\end{tabular}%
\caption{Normalized Mean Absolute Deviation $\left\langle |\E n - n| / n\right\rangle $.}%
$\label{tbl:intlErrors}$%
\end{table}%
\end{center}

As can be seen from the above results, the probabilistic model of the number
of relevant parties is an improvement over earlier models like the
Seat--Product Model, scoring better for eight of ten tested countries, and
worse only for Wallonia (where the whole sample consists of a single
election) and Netherlands (where the ease of registering small parties makes
the uniform distribution assumption probably unrealistic).

\subsection{Polish County Elections}

There are currently 314 counties in Poland. In each county, there is a
council (varying between 15 and 60 seats) elected using the
Jefferson-D'Hondt with the statutory threshold of 5\%. Each county is
divided into three or more districts with magnitude greater than or equal to 
$3$. For each county, we analyze all elections between 1998 and 2014
(yielding 1564 elections), again computing the expected number of relevant
parties using our probabilistic model and comparing the outcome with the one
arising under the Seat-Product Model. Mean error values are given in Table %
\ref{tbl:cntyErrors}.

\begin{center}
\begin{table}[h] \centering%
\begin{tabular}{c|cc}
& mean abs. dev. & mean square error \\ 
model & $\left\langle \left\vert \func{E}n-n\right\vert \right\rangle $ & $%
\sqrt{\left\langle \left( \func{E}n-n\right) ^{2}\right\rangle }$ \\ 
\hline\hline
probabilistic & 0.945 & 1.149 \\ 
Seat-Product Model & 2.068 & 2.460 \\ 
$n=N$ & 2.111 & 2.466%
\end{tabular}%
\caption{Overall Errors of the Three Alternative Models of the Number of Relevant Parties.}%
$\label{tbl:cntyErrors}$%
\end{table}%
\end{center}

\begin{figure}[h]
	\centering
	\includegraphics[width=0.6\linewidth]{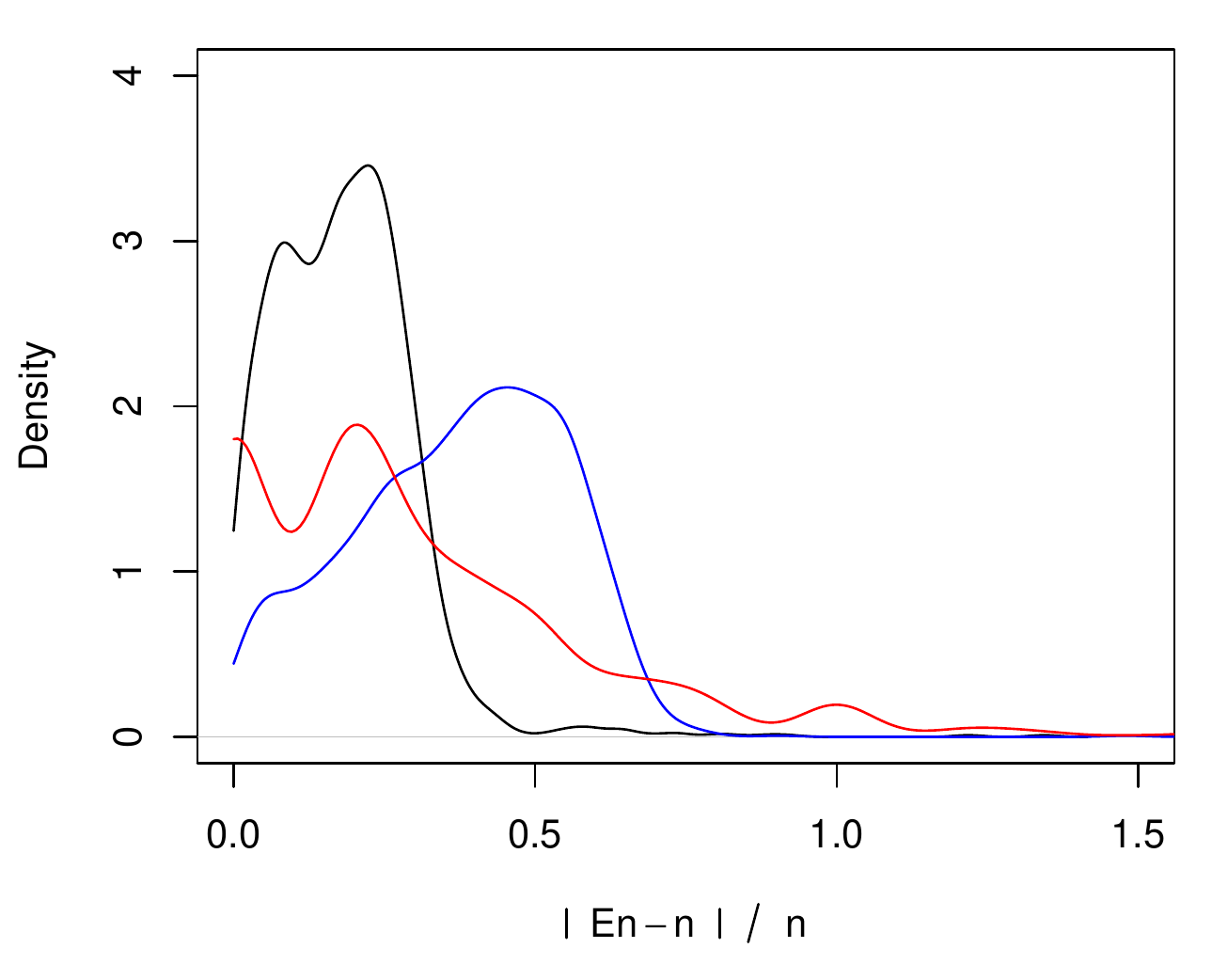}
	\caption{Kernel density estimates of the normalized absolute deviation $|\E n - n| / n$ for the
		probabilistic model (black curve), the Seat-Product Model (blue curve) and
		the ``naive'' ($n = N$) model (red curve).}
	\label{fig:densErrIntl}
\end{figure}%

Again, the probabilistic model represents a significant improvement over the
alternatives.%
\FloatBarrier%

\bibliographystyle{authordate1}
\bibliography{acompat,expectNoParties}

\appendix
\section{Distribution of $P_{k}^{\downarrow }$ and $%
V_{k}^{\downarrow }$}

Let $X_{1},\dots ,X_{N}$ be independent random variables distributed
according to $\func{Exp}\left( 1\right) $, i.e., with density $f\left(
x\right) :=e^{-x}$. It is well known (see, e.g, 
\citealp{Jambunathan54}%
) that%
\begin{equation}
\left( \frac{X_{1}}{\sum_{i=1}^{N}X_{i}},\dots ,\frac{X_{N}}{%
\sum_{i=1}^{N}X_{i}}\right) \sim \func{Unif}\left( \Delta _{N}\right) .
\end{equation}

\subsection{$k$-th order statistic $X_{k}^{\downarrow }$}

Consider first the density of $X_{k}^{\downarrow }$. We start with the
following result by 
\citet{Renyi53}%
:%
\begin{equation}
X_{k}^{\downarrow }\overset{d}{=}\sum_{j=k}^{N}\frac{Z_{j}}{j},
\end{equation}%
where $Z_{j},\dots ,Z_{N}\sim \func{Exp}\left( 1\right) $ and are i.i.d.
Thus,%
\begin{equation}
\varphi _{X_{k}^{\downarrow }}(s)=\prod\limits_{j=k}^{N}\left( 1-\frac{is}{j}%
\right) ^{-1}=\prod\limits_{j=k}^{N}\frac{j}{j-is}=\frac{\left( k\right)
_{N-k+1}}{\left( k-is\right) _{N-k+1}},  \label{characterXk}
\end{equation}%
wherefore (by L\'{e}vy's inversion formula, 
\citealp[p.~347, (26.20)]{Billingsley95}%
)%
\begin{equation}
f_{X_{k}^{\downarrow }}(x)=\frac{\left( k\right) _{N-k+1}}{2\pi }%
\int_{-\infty }^{+\infty }\frac{e^{-isx}}{\left( k-is\right) _{N-k+1}}ds.
\end{equation}%
Let us denote the function under the integral as $\lambda \left( s\right) $.
Note that it has $N-k+1$ simple poles at $-ik,\dots ,-iN$.

Let us consider a contour $C_{r}:=\left[ r,-r\right] \cup A_{r}$, where%
\begin{equation}
A_{r}:=\left\{ x\in \mathbb{C}:\left( \left\vert x\right\vert =r\right) 
\text{ and }\arg (x)\in (\pi ,2\pi )\right\} 
\end{equation}%
is a positively oriented arc from $-r$ to $r$ centered at $0$, and $r>N$.
Then by the residue theorem,%
\begin{equation}
\oint\nolimits_{C_{r}}\lambda \left( s\right) \,ds=2\pi
i\,\sum\nolimits_{j=k}^{N}\func{Res}\left( \lambda ,-ij\right) .
\end{equation}%
Let us first consider the integral of $\lambda \left( s\right) $ over $A_{r}$%
. Let us substitute $s=re^{i\theta }$, where $\theta \in \left( \pi ,2\pi
\right) $. Then for $s\in A_{r}$ we have%
\begin{eqnarray}
\left\vert \lambda \left( s\right) \right\vert  &\leq &e^{rx\sin \theta
}\prod\limits_{j=k}^{N}\left\vert \left( j+r\sin \theta -ir\cos \theta
\right) ^{-1}\right\vert = \\
&=&e^{rx\sin \theta }\prod\limits_{j=k}^{N}\left( j^{2}+r^{2}+2jr\sin \theta
\right) ^{-1},
\end{eqnarray}%
which is dominated by $e^{rx\sin \theta }r^{-2(N-k+1)}$ as $r\rightarrow
\infty $. Therefore by the estimation lemma 
\citep[Theorem~5.24]{Howie03}%
,%
\begin{equation}
\oint\nolimits_{A_{r}}\lambda \left( s\right) \,ds=O\left( e^{rx\sin \theta
}r^{-2(N-k+1)}r\right) =O\left( 0\right) 
\end{equation}%
as $r\rightarrow \infty $, wherefore%
\begin{equation}
\int_{-\infty }^{+\infty }\lambda \left( s\right) \,ds=-2\pi
i\,\sum\nolimits_{j=k}^{N}\func{Res}\left( \lambda ,-ij\right) 
\end{equation}%
(the minus sign in the integral arises because the contour integral is
computed in the reverse direction). Computing the residues, we get%
\begin{eqnarray}
\func{Res}\left( \lambda ,-ij\right)  &=&\underset{s\rightarrow -ij}{\lim }%
\left( s+ij\right) e^{-isx}\prod\nolimits_{l=k}^{N}\frac{1}{l-is}= \\
&=&ie^{-jx}\prod\nolimits_{l=k}^{j-1}\frac{1}{l-j}\prod\nolimits_{l=j+1}^{N}%
\frac{1}{l-j}= \\
&=&i\left( -1\right) ^{k-j}\frac{1}{\left( j-k\right) !}\frac{1}{\left(
N-j\right) !}e^{-jx},
\end{eqnarray}%
and thus%
\begin{gather}
\int_{-\infty }^{+\infty }\lambda \left( s\right) \,ds=-2\pi
i\,\sum\nolimits_{j=k}^{N}i\left( -1\right) ^{k-j}\frac{1}{\left( j-k\right)
!}\frac{1}{\left( N-j\right) !}e^{-jx}= \\
=2\pi \frac{e^{-kx}\left( 1-e^{-x}\right) ^{N-k}}{\left( N-k\right) !}.
\end{gather}%
Substituting the above into (\ref{characterXk}), we obtain the density of $%
X_{k}^{\downarrow }$: 
\begin{equation}
f_{X_{k}^{\downarrow }}(x)=\frac{N!}{\left( k-1\right) !}\frac{e^{-kx}\left(
1-e^{-x}\right) ^{N-k}}{\left( N-k\right) !}=k\dbinom{N}{k}e^{-kx}\left(
1-e^{-x}\right) ^{N-k}.
\end{equation}

\subsection{Vote shares}

Let us fix the $k$-th largest order statistic, $X_{k}^{\downarrow }$, at $y$%
, and let:

\begin{itemize}
\item $\Psi ^{\downarrow }:=\sum\nolimits_{j=k+1}^{N}X_{j}^{\downarrow },$

\item $\Psi ^{\uparrow }:=\sum\nolimits_{j=1}^{k-1}X_{j}^{\downarrow },$

\item $\Psi :=\sum\nolimits_{j=1}^{N}X_{j}^{\downarrow }=\Psi ^{\downarrow
}+\Psi ^{\uparrow }+X_{k}^{\downarrow }.$
\end{itemize}

The non-normalized vote share of the $k$-th party is given by%
\begin{equation}
V_{k}^{\downarrow }=\frac{X_{k}^{\downarrow }}{\Psi }.
\end{equation}

The normalized vote share of the $k$-th party assuming $k$ relevant parties
is given by%
\begin{equation}
P_{k}^{\downarrow }=\frac{V_{k}^{\downarrow }}{\sum_{j=1}^{k}V_{j}^{%
\downarrow }}=\frac{V_{k}^{\downarrow }}{\Psi }\frac{\Psi }{%
\sum_{j=1}^{k}V_{k}^{\downarrow }}=\frac{V_{k}^{\downarrow }}{%
V_{k}^{\downarrow }+\Psi ^{\uparrow }}.
\end{equation}

Note that $\Psi ^{\uparrow }=\sum\nolimits_{j=1}^{k-1}Y_{j}$, where $%
Y_{j}\sim \func{Exp}\left( 1\right) $ truncated to $(y,\infty )$, and $\Psi
^{\downarrow }=\sum\nolimits_{j=1}^{N-k}Z_{j}$, where $Z_{j}\sim \func{Exp}%
\left( 1\right) $ truncated to $(0,y)$. Then the density of $Z_{j}$ is given
by%
\begin{equation}
f_{Z}(x)=\frac{e^{-x}}{F(y)-F(0)}=\frac{e^{-x}}{1-e^{-y}}
\end{equation}%
for $x\in (0,y)$, and the density of $Y_{j}$ is given by%
\begin{equation}
f_{Y}(x)=\frac{e^{-x}}{1-F(y)}=e^{y-x}
\end{equation}%
for $x\in (y,\infty )$, and their characteristic functions are given by,
respectively,%
\begin{equation}
\varphi _{Z}\left( t\right) =\frac{1}{1-e^{-y}}\int_{0}^{y}e^{itx-x}dx=\frac{%
i}{i+t}\frac{e^{y}-e^{ity}}{e^{y}-1},
\end{equation}

and%
\begin{equation}
\varphi _{Y}\left( t\right) =e^{y}\int_{y}^{\infty }e^{itx-x}dx=\frac{i}{i+t}%
e^{ity}.
\end{equation}%
Hence, the conditional characteristic function of $\Psi ^{\downarrow }$ is
given by%
\begin{equation}
\varphi _{\Psi ^{\downarrow }|X_{k}^{\downarrow }}(t,y)=\left( \frac{i}{i+t}%
\frac{e^{y}-e^{ity}}{e^{y}-1}\right) ^{N-k}=\left( 1-it\right) ^{k-N}\left( 
\frac{e^{y}-e^{ity}}{e^{y}-1}\right) ^{N-k},
\end{equation}%
the conditional characteristic function of $\Psi ^{\uparrow }$ by%
\begin{equation}
\varphi _{\Psi ^{\uparrow }|X_{k}^{\downarrow }}(t,y)=\left( \frac{i}{i+t}%
e^{ity}\right) ^{k-1}=\left( 1-it\right) ^{1-k}e^{ity\left( k-1\right) },
\end{equation}%
and the conditional characteristic function of $\Psi $ by 
\begin{eqnarray}
\varphi _{\Psi |X_{k}^{\downarrow }}(t,y) &=&e^{ity}\left( 1-it\right)
^{1-N}\left( \frac{e^{y}-e^{ity}}{e^{y}-1}\right) ^{N-k}e^{ity\left(
k-1\right) }= \\
&=&\left( 1-it\right) ^{1-N}\left( \frac{e^{y}-e^{ity}}{e^{y}-1}\right)
^{N-k}e^{ityk}.
\end{eqnarray}%
By the binomial theorem,%
\begin{equation}
\left( e^{y}-e^{ity}\right) ^{N-k}=\sum\nolimits_{a=0}^{N-k}\dbinom{N-k}{a}%
e^{y(N-k-a)}e^{i(ty+\pi )a},
\end{equation}%
wherefore%
\begin{equation}
\varphi _{\Psi ^{\downarrow }|X_{k}^{\downarrow
}}(t,y)=(e^{y}-1)^{k-N}\sum\nolimits_{a=0}^{N-k}\dbinom{N-k}{a}\left(
1-it\right) ^{k-N}e^{y(N-k-a)}e^{i\pi a}e^{itya}.
\end{equation}

Again by Levy's inversion formula,%
\begin{gather}
f_{\Psi ^{\downarrow }|X_{k}^{\downarrow }}(x,y)=\frac{1}{2\pi }%
(e^{y}-1)^{k-N}\int_{-\infty }^{+\infty }\sum\nolimits_{a=0}^{N-k}\dbinom{N-k%
}{a}\left( 1-it\right) ^{k-N}e^{y(N-k-a)}e^{i\pi a}e^{itya-itx}\,dt= \\
=\frac{1}{2\pi }(e^{y}-1)^{k-N}\sum\nolimits_{a=0}^{N-k}\dbinom{N-k}{a}%
e^{y(N-k-a)}i^{2a}\int_{-\infty }^{+\infty }\left( 1-it\right)
^{k-N}e^{it(ya-x)}\,dt= \\
=\frac{1}{2\pi }(e^{y}-1)^{k-N}\sum\nolimits_{a=0}^{N-k}\left( -1\right) ^{a}%
\dbinom{N-k}{a}e^{y(N-k-a)}\tciFourier \left\{ \left( 1-it\right)
^{k-N}\right\} (x-ya)= \\
=\frac{1}{2\pi }(e^{y}-1)^{k-N}\sum\nolimits_{a=k}^{N}\left( -1\right) ^{a-k}%
\dbinom{N-k}{a-k}e^{y(N-a)}\tciFourier \left\{ \left( 1-it\right)
^{-(N-k)}\right\} (x-y(a-k)).
\end{gather}%
Now by 
\citealp[\S~3.2~(3), p.~118]{Bateman54}%
,%
\begin{gather}
f_{\Psi ^{\downarrow }|X_{k}^{\downarrow }}(x,y)= \\
=\frac{(e^{y}-1)^{k-N}}{2\pi }\sum_{a=k}^{N}\left( -1\right) ^{a-k}%
\dbinom{N-k}{a-k}e^{y(N-a)}\left\{ 
\begin{array}{cc}
\frac{2\pi \left( x-y(a-k)\right) ^{N-k-1}e^{-(x-y(a-k))}}{\Gamma \left(
N-k\right) } & :x>y(a-k) \\ 
0 & :x\leq y(a-k)%
\end{array}%
\right. = \\
=\frac{(1-e^{-y})^{k-N}}{\left( N-k-1\right) !}e^{-x}\sum\nolimits_{a=k}^{%
\min \left( N,\left\lfloor x/y\right\rfloor +k\right) }\left( -1\right)
^{a-k}\dbinom{N-k}{a-k}\left( x-y(a-k)\right) ^{N-k-1}.
\end{gather}%
Likewise,%
\begin{eqnarray}
f_{\Psi ^{\uparrow }|X_{k}^{\downarrow }}(x,y) &=&\frac{1}{2\pi }%
\int_{-\infty }^{+\infty }\left( 1-it\right) ^{1-k}e^{ity(k-1)-itx}\,dt= \\
&=&\frac{1}{2\pi }\mathcal{\tciFourier }\left\{ \left( 1-it\right)
^{1-k}\right\} (x-y(k-1))= \\
&=&\frac{1}{\left( k-2\right) !}\left\{ 
\begin{array}{cc}
\left( x-y(k-1)\right) ^{k-2}e^{y(k-1)-x} & :x>y(k-1) \\ 
0 & :x\leq y(k-1)%
\end{array}%
\right. ,
\end{eqnarray}%
and%
\begin{equation}
f_{\Psi ^{\uparrow }+X_{k}^{\downarrow }|X_{k}^{\downarrow }}(x,y)=f_{\Psi
^{\uparrow }|X_{k}^{\downarrow }}(x-y,y)=\frac{1}{\left( k-2\right) !}%
\left\{ 
\begin{array}{cc}
\left( x-yk\right) ^{k-2}e^{yk-x} & :x>yk \\ 
0 & :x\leq yk%
\end{array}%
\right. .
\end{equation}%

\subsection{Joint density and ratio}

Joint density is easy to obtain from the above results:%
\begin{eqnarray}
f_{\Psi ^{\uparrow }+X_{k}^{\downarrow },X_{k}^{\downarrow }}(x,y) &=&\frac{1%
}{\left( k-2\right) !}\left\{ 
\begin{array}{cc}
\left( x-yk\right) ^{k-2}e^{yk-x} & :x>yk \\ 
0 & :x\leq yk%
\end{array}%
\right. \times k\dbinom{N}{k}e^{-ky}\left( 1-e^{-y}\right) ^{N-k}= \\
&=&\mathbf{1}_{\left\{ x>yk\right\} }\frac{k}{\left( k-2\right) !}\dbinom{N}{%
k}\left( 1-e^{-y}\right) ^{N-k}\left( x-yk\right) ^{k-2}e^{-x}.
\end{eqnarray}%
And the density of the ratio $P_{k}^{\downarrow }=X_{k}^{\downarrow }/\left(
X_{k}^{\downarrow }+\Psi ^{\uparrow }\right) $ is given by%
\begin{eqnarray}
f_{P_{k}^{\downarrow }}\left( z\right) &=&\int_{0}^{\infty }x\,\mathbf{1}%
_{\left\{ x>zk\right\} }\frac{k}{\left( k-2\right) !}\dbinom{N}{k}\left(
1-e^{-zx}\right) ^{N-k}\left( x-zxk\right) ^{k-2}e^{-x}\,dx= \\
&=&\mathbf{1}_{\left\{ z<1/k\right\} }\frac{k}{\left( k-2\right) !}\dbinom{N%
}{k}\left( 1-zk\right) ^{k-2}\int_{0}^{\infty }x^{k-1}e^{-x}\,\left(
1-e^{-zx}\right) ^{N-k}\,dx= \\
&=&\mathbf{1}_{\left\{ z<1/k\right\} }\frac{k}{\left( k-2\right) !}\dbinom{N%
}{k}\left( 1-zk\right) ^{k-2}\int_{0}^{\infty
}x^{k-1}e^{-x}\,\sum_{a=0}^{N-k}\left( -1\right) ^{a}\dbinom{N-k}{a}%
e^{-zxa}\,dx= \\
&=&\mathbf{1}_{\left\{ z<1/k\right\} }\frac{k}{\left( k-2\right) !}\dbinom{N%
}{k}\left( 1-zk\right) ^{k-2}\sum_{a=0}^{N-k}\left( -1\right) ^{a}\dbinom{N-k%
}{a}\int_{0}^{\infty }x^{k-1}e^{-zxa-x}\,dx= \\
&=&\mathbf{1}_{\left\{ z<1/k\right\} }\frac{k}{\left( k-2\right) !}\dbinom{N%
}{k}\left( 1-zk\right) ^{k-2}\sum_{a=0}^{N-k}\left( -1\right) ^{a}\dbinom{N-k%
}{a}\left( 1+az\right) ^{-k}\Gamma \left( k\right) = \\
&=&\mathbf{1}_{\left\{ z<1/k\right\} }\,k\left( k-1\right) \dbinom{N}{k}%
\left( 1-zk\right) ^{k-2}\sum_{a=0}^{N-k}\left( -1\right) ^{a}\dbinom{N-k}{a}%
\left( 1+az\right) ^{-k}.
\end{eqnarray}%

\section{Obtaining the probability of relevance}

We are interested in $\Pr \left( P_{k}^{\downarrow }>t_{k}\right) $. By
integrating the density of $P_{k}^{\downarrow }$, we obtain%
\begin{gather}
\Pr \left( P_{k}^{\downarrow }>t_{k}\right) =\int_{\left( 2m+k\right)
^{-1}}^{k^{-1}}f_{P_{k}^{\downarrow }}\left( x\right) \,dx= \\
=k\left( k-1\right) \dbinom{N}{k}\sum_{a=0}^{N-k}\left( -1\right) ^{a}%
\dbinom{N-k}{a}\int_{\left( 2m+k\right) ^{-1}}^{k^{-1}}\frac{\left(
1-xk\right) ^{k-2}}{\left( 1+ax\right) ^{k}}\,dx= \\
=k\left( k-1\right) \dbinom{N}{k}\sum_{a=0}^{N-k}\left( -1\right) ^{a}%
\dbinom{N-k}{a}\left. \frac{-1}{\left( k-1\right) \left( a+k\right) }\left( 
\frac{1-xk}{1+ax}\right) ^{k-1}\right\vert _{x=\left( 2m+k\right)
^{-1}}^{x=k^{-1}}= \\
=k\left( k-1\right) \dbinom{N}{k}\sum_{a=0}^{N-k}\left( -1\right) ^{a}%
\dbinom{N-k}{a}\frac{1}{\left( k-1\right) \left( a+k\right) }\left( \frac{2m%
}{a+k+2m}\right) ^{k-1}= \\
=k\dbinom{N}{k}\sum_{a=k}^{N}\left( -1\right) ^{a-k}\dbinom{N-k}{a-k}\frac{1%
}{a}\left( \frac{2m}{a+2m}\right) ^{k-1}.
\end{gather}

Likewise, we are interested in $\Pr \left( V_{k}^{\downarrow }>\tau \right) $%
. By integrating the density of $V_{k}^{\downarrow }$, we obtain 
\begin{gather}
\Pr \left( V_{k}^{\downarrow }>\tau \right) =\int_{\tau
}^{k^{-1}}f_{V_{k}^{\downarrow }}\left( x\right) \,dx= \\
=N\left( N-1\right) \dbinom{N-1}{k-1}\int_{\tau
}^{k^{-1}}\sum\limits_{a=k}^{\min \left( N,\left\lfloor 1/x\right\rfloor
\right) }\left( -1\right) ^{a-k}\dbinom{N-k}{a-k}\left( 1-ax\right)
^{N-2}\,dx= \\
=N\left( N-1\right) \dbinom{N-1}{k-1}\sum_{a=k}^{\min \left( N,\left\lfloor
\tau ^{-1}\right\rfloor \right) }\left( -1\right) ^{a-k}\dbinom{N-k}{a-k}%
\int_{\tau }^{a^{-1}}\left( 1-ax\right) ^{N-2}\,dx= \\
=N\left( N-1\right) \dbinom{N-1}{k-1}\sum_{a=k}^{\min \left( N,\left\lfloor
\tau ^{-1}\right\rfloor \right) }\left( -1\right) ^{a-k}\dbinom{N-k}{a-k}%
\frac{\left( 1-a\tau \right) ^{N-1}}{a\left( N-1\right) }= \\
=N\dbinom{N-1}{k-1}\sum_{a=k}^{\min \left( N,\left\lfloor \tau
^{-1}\right\rfloor \right) }\left( -1\right) ^{a-k}\dbinom{N-k}{a-k}\frac{1}{%
a}\left( 1-a\tau \right) ^{N-1}.
\end{gather}

\section{Obtaining the moments of the distribution of the number of relevant parties}

First moment (expectation):
\begin{gather}
\func{E}n=\sum_{k=1}^{N}k\Pr \left( n=k\right) =\sum_{k=1}^{N}k\Pr \left(
P_{k}^{\downarrow }>t_{k}\right) -\sum_{k=1}^{N-1}k\Pr \left(
P_{k+1}^{\downarrow }>t_{k+1}\right) = \\
=\sum_{k=1}^{N}k\Pr \left( P_{k}^{\downarrow }>t_{k}\right)
-\sum_{k=2}^{N}\left( k-1\right) \Pr \left( P_{k}^{\downarrow }>t_{k}\right)
= \\
=\sum_{k=1}^{N}k\Pr \left( P_{k}^{\downarrow }>t_{k}\right)
-\sum_{k=2}^{N}k\Pr \left( P_{k}^{\downarrow }>t_{k}\right)
+\sum_{k=2}^{N}\Pr \left( P_{k}^{\downarrow }>t_{k}\right) =
\sum_{k=1}^{N}\Pr \left( P_{k}^{\downarrow }>t_{k}\right) .
\end{gather}

Second moment:%
\begin{gather}
\func{E}\left( n^{2}\right) =\sum_{k=1}^{N}k^{2}\Pr \left( n=k\right)
=\sum_{k=1}^{N}k^{2}\Pr \left( P_{k}^{\downarrow }>t_{k}\right)
-\sum_{k=1}^{N-1}k^{2}\Pr \left( P_{k+1}^{\downarrow }>t_{k+1}\right) = \\
=\sum_{k=1}^{N}k^{2}\Pr \left( P_{k}^{\downarrow }>t_{k}\right)
-\sum_{k=2}^{N}\left( k-1\right) ^{2}\Pr \left( P_{k}^{\downarrow
}>t_{k}\right) = \\
=\Pr \left( P_{1}^{\downarrow }>t_{1}\right) -\sum_{k=2}^{N}\Pr \left(
P_{k}^{\downarrow }>t_{k}\right) +2\sum_{k=2}^{N}k\Pr \left(
P_{k}^{\downarrow }>t_{k}\right) = \\
=2\sum_{k=2}^{N}k\Pr \left( P_{k}^{\downarrow }>t_{k}\right)
-\sum_{k=1}^{N}\Pr \left( P_{k}^{\downarrow }>t_{k}\right) .
\end{gather}

\end{document}